\documentclass[
    ,final            
  ,numberedheadings 
  ]
  {aipproc}

\layoutstyle{8x11double}

\usepackage{mathptmx}
\usepackage[latin1]{inputenc}
\usepackage{units}
\usepackage[final]{graphicx}
\usepackage{amsxtra,amsfonts,amssymb,amsmath,fancybox,float}
\usepackage{fancyhdr, xspace}
\usepackage[english]{babel}

\begin{document}

\title{Dark Matter Searches with the Fermi Large Area Telescope}



\author{Christine Meurer on behalf of the Fermi LAT collaboration}{
  address={Fysikum, Stockholms Universitet, AlbaNova Universitetscentrum, 106 91  Stockholm}
}

\begin{abstract}
The Fermi Gamma-Ray Space Telescope, successfully launched on 
June~11th,~2008, is the next generation satellite experiment for high-energy gamma-ray 
astronomy. The main instrument, the Fermi Large Area Telescope (LAT), with a wide field 
of view (>~\unit[2]{sr}), a large effective area (>~\unit[8000]{$\text{cm}^{2}$} at 
\unit[1]{GeV}), sub-arcminute source localization, a large energy range (\unit[20]{MeV} 
- \unit[300]{GeV}) and a good energy resolution (close to 8\% at \unit[1]{GeV}), has 
excellent potential to either discover or to constrain a Dark Matter signal.
The Fermi LAT team pursues complementary searches for signatures of particle Dark Matter 
in different search regions such as the galactic center, galactic satellites and subhalos, 
the milky way halo, extragalactic regions as well as the search for spectral lines.
In these proceedings we examine the potential of the LAT to detect gamma-rays coming from
Weakly Interacting Massive Particle annihilations in these regions with special 
focus on the galactic center region.
\end{abstract}

\maketitle


\section{Introduction}
The Fermi Gamma-Ray Space Telescope (in short ``Fermi'')~\cite{Atwood:1993zn, Michelson:2007zz}, which is part of the NASA's office of Space and Science strategic plan, is an international space mission that studies cosmic $\gamma$-rays in the energy range \unit[20]{MeV} - \unit[300]{GeV}.  This mission is realized, as a close collaboration between the astrophysics and particle physics communities, including institutions in the USA, Japan, France, Germany, Italy and Sweden. The main instrument on Fermi is the Large Area Telescope (LAT) complemented by a dedicated instrument for the detection of gamma-ray bursts (the Gamma-ray burst monitor, GBM).

The main science targets of the LAT detector are 
\begin{itemize}
\item to understand the mechanisms of particle acceleration in active galactic nuclei, pulsars, and supernova remnants 
\item to resolve the gamma-ray sky; unidentified sources and diffuse emission 
\item determine the high-energy behavior of gamma-ray bursts and transients  
\item to probe Dark Matter and early Universe
\end{itemize}

\begin{figure}[h]
 \includegraphics[width=0.38\textwidth, viewport= 120 100 820 1000,clip]{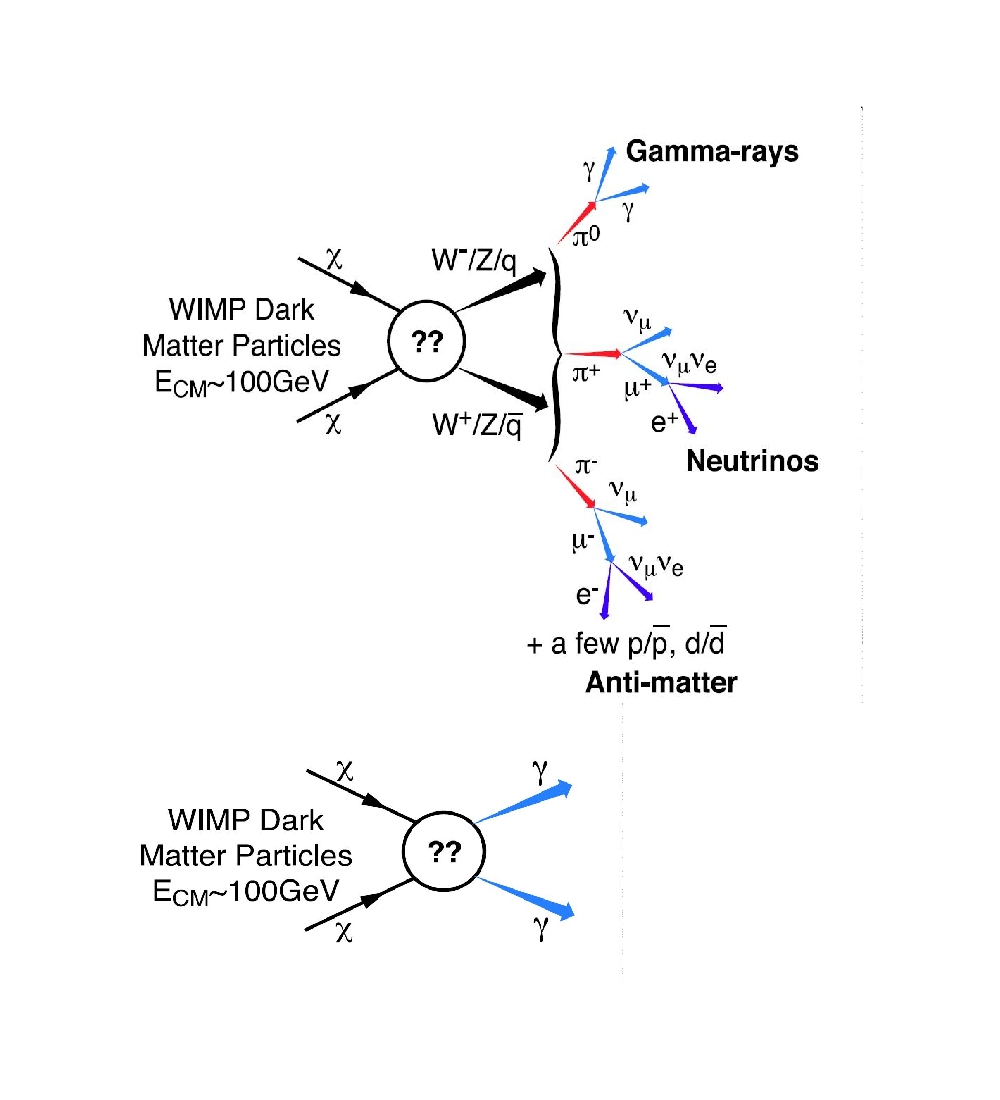}
  \caption{Sketch of the annihilation of two WIMPs. \hspace{0.9cm}
Upper plot: Dominant annihilation channel into a quark-antiquark pair (or two massive gauge bosons),whereas in the resulting hadronization process $\pi^0$s are produced which decay into two $\gamma$s and lead to a continuum $\gamma$-ray spectrum. 
Lower plot: Suppressed annihilation channel into two $\gamma$s which leads to a line signal in the $\gamma$-ray spectrum. \label{WIMPdecay}}
\end{figure}

There is compelling experimental evidence for a dark component of the matter density of the Universe from observation of on many different scales such as galaxies, galaxy clusters and cosmic background radiation~\cite{Bertone:2004pz}. 
The questions of what constitutes this Dark Matter is one of the great mysteries of modern physics. 
One of the most promising candidates for the Dark Matter is a Weakly Interacting Massive particle (WIMP). As predicted in many extensions of the Standard Model of Particle Physics, 
WIMPs (e.g. neutralinos) can be detected indirectly via their annihilation products, in particular neutrinos, anti-protons, positrons and gamma-rays (see Fig.~\ref{WIMPdecay}). These annihilations of WIMPs may give rise to a signal in gamma-ray spectra from many cosmic sources.

\section{Gamma Ray Signal from Dark Matter}

The spectrum of gamma-rays due to WIMP annihilation can be constructed in a ``generic'' fashion, i.e. almost independent of underlying physics model.  
The $\gamma$-ray continuum flux from WIMP annihilation at a given photon energy $E_\gamma$ from a direction that forms an angle $\psi$ between the direction of the galactic center and that
of observation is given by \cite{Gondolo:2004sc}:

\begin{eqnarray*}
\phi_{\text{WIMP}}(E_{\gamma},\psi) = \frac{1}{4\pi}\frac{<\sigma_\text{ann}v>}{2m^2_\text{WIMP}} 
\sum{\frac{dN_\gamma^f}{dE_\gamma}B_f } \int_\text{los}\rho(l)^2 dl(\psi) 
\label{eq:yield}
\end{eqnarray*}

The particle physics model enters through the WIMP mass $m_\text{WIMP}$ , the total mean annihilation cross-section $\sigma_\text{ann}$ multiplied by the relative velocity of the particles (in the limit of $v \rightarrow 0$), and the sum of all the photon yields $dN_\gamma^f/dE_\gamma$ for each annihilation channel weighted by the corresponding branching ratio $B_f$ .
For Majorana fermion WIMPs light fermions are suppressed so that the dominant fermionic annihilation products will be $b\bar{b},t\bar{t}$ and $\tau^+\tau^-$. Apart from the $\tau^+\tau^-$ channel, the photon yields are quite similar \cite{Cesarini:2003nr}. The continuum $\gamma$-ray spectrum shows a cutoff at the energy of the corresponding WIMP mass.
In the special case of the 2 $\gamma$ final state a spectral line is centered on energy $E_\gamma=m_\text{WIMP}$. This process is loop suppressed with a branching fraction of usually $10^{-3}$ to $10^{-4}$. However, see also~\cite{Gustafsson:2007pc}.
The spatial distribution of the WIMPs is considered in the integral along the line of sight (los) of the assumed density squared $\rho(l)^2$.
We simulate the photon yield by using the Monte Carlo program Pythia~6.202 \cite{Sjostrand:1993yb}, which is implemented in the DarkSUSY package \cite{Gondolo:2004sc}. For the density distribution of the WIMPs a Navarro-Frenk-White (NFW) profile is assumed \cite{Navarro:1995iw}. 
For details of this simulation see \cite{Baltz:2008wd}.

\section{LAT performance}
The baseline of the Fermi LAT detector is modular, consisting of a 4~$\times$~4 array of identical towers. Each 40~$\times$~\unit[40]{cm$^2$} tower comprises a tracker, calorimeter and data acquisition module. The tracking detector consists of 18~xy~layers of silicon strip detectors. This detector technology has a long and successful history of application in accelerator-based high-energy physics. It is well-matched to the requirements of high detection efficiency (\unit[$>$ 99]{\%}), excellent position resolution (\unit[$<$~60]{$\mu$m}), large signal/noise ($>$~20), negligible cross-talk, and ease of trigger and readout.  Compared to its predecessor EGRET~\cite{EGRET}, the LAT has a sensitivity exceeding that of EGRET by at least a factor of 30, the energy range is extended by a factor 10 and the energy (Fermi: \unit[8]{\%} at \unit[10]{GeV}) and angular resolutions (Fermi $\text{PSF}_{68\%} = 0.1^{\circ}$ at \unit[10]{GeV}) are improved by a factor of at least two. The improvement in sensitivity is partly due to the choice of silicon tracking detectors instead of the spark-chambers used in EGRET, which reduces the dead-time by more than three orders of magnitude.
 Fermi was successfully launched on June 11th, 2008 and after 60 days of commissioning it delivers continuously data. 

\section{Complementary Searches for Dark Matter with Fermi}
The Fermi LAT collaboration pursues complementary searches for Dark Matter each presenting its own challenges and advantages. In Tab.~\ref{tab1} we summarize the most important ones. In the following we focus on the discussion on the Dark Matter analysis technique in the region of the galactic center.

\begin{table}[h]
\begin{tabular}{lll}
\hline
    \tablehead{1}{l}{b}{Search \\ technique}
  & \tablehead{1}{l}{b}{Advantages}
  & \tablehead{1}{l}{b}{Challenges}   \\
\hline
Galactic Center & good statistics & source confusion, \\
                &                 & galactic diffuse background \\
\hline
Galactic Halo   & very good statistics & galactic diffuse background \\
\hline
Satellites   & low background, & low statistics \\
                & good source id & \\  
\hline
Spectral Lines  & no astrophysical uncertainties, & low statistics \\ 
                & good source id & \\ 
\hline
Extragalactic   & very good statistics & astrophysics, \\ 
                &                      & galactic diffuse background \\
\hline
\end{tabular}
\caption{Overview of the various sources for particle Dark Matter undertaken by the Fermi LAT collaboration. For more details see \cite{Baltz:2008wd}. \label{tab1}}
\end{table}

\subsection{The Galactic Center}
The center of the Milky Way is a formidable astrophysical target to search for a Dark Matter signal, due to its coincidence with the cuspid part of the Dark Matter halo density profile and since the WIMP annihilation rate is proportional to the density squared, significant fluxes can be expected.
On the other hand, establishing a signal requires identification of the high energy gamma-ray sources which are close (or near) the center~\cite{Mayer-Hasselwander:1998hg} and also an adequate modeling of the galactic diffuse emission due to cosmic rays colliding with the interstellar medium. 

\begin{figure}[h]
  \includegraphics[width=.45\textwidth]{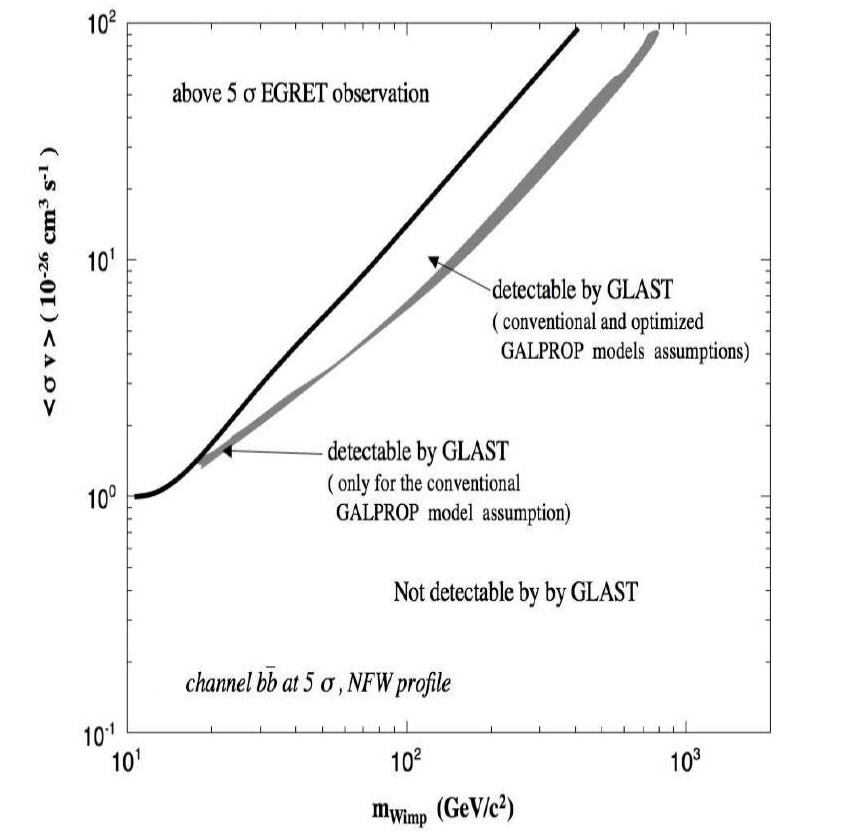}
  \caption{Fermi sensitivity for a $b\bar{b}$ annihilation signal in the $\gamma$-ray spectrum from the galactic center region in terms of the velocity averaged cross-section versus the WIMP mass. The upper part of the phasespace is already excluded by the EGRET data and the lower part is not detectable by Fermi~\cite{Baltz:2008wd}.  \label{GCbb}}
\end{figure}

In a first step we perform an analysis, which is based on the hypothesis that the main task will be to distinguish the Dark Matter signal from a galactic diffuse background after the astrophysical sources are disentangled and subtracted using the information provided by spectral and angular analysis and multiwavelength observations. The systematic uncertainty in the assessment of the diffuse flux caused by this subtraction is neglected here.
Furthermore, only one dominant annihilation channel (e.g. $b\bar{b}$) is considered at a time. We generate the WIMP annihilation differential fluxes above \unit[1]{GeV} and in a region of \unit[0.5]{$^\circ$} radius around the galactic center. 
The EGRET flux modeled in \cite{Mayer-Hasselwander:1998hg} is also simulated at the galactic center and a standard $\chi^2$ statistical analysis is performed to check if a given WIMP model conflicts with EGRET data at the \unit[5]{$\sigma$} level. For models compatible with EGRET data, a second $\chi^2$ test is performed to check if Fermi is able to disentangle the WIMP contribution from either the {\it conventional}~\cite{Strong:1998fr} or the {\it optimized}~\cite{Strong:2004de} GALPROP galactic diffuse background model. 
Fig.~\ref{GCbb} shows the result of the scan at a \unit[5]{$\sigma$} confidence level for the $b\bar{b}$ final state for 5 years of all sky scanning operation. It can be seen that regions of \unit[$\mathcal{O}(10^{-26})$]{$\text{cm}^3\text{s}^{-1}$} are within the reach of Fermi.

In a next step, prior to the Fermi launch, we prepared analyses methods using a realistic one year full sky simulation, the ``GLAST Service Challenge data 2'' (\mbox{SCdata2}), consisting of a realistic sky (around 20,000 sources, including a continuum Dark Matter component modeled with DarkSUSY assuming a WIMP mass of \unit[107.9]{GeV} and a spatial distribution following a NFW profile with the center at the galactic center). 

To establish a discovery of Dark Matter in the region of the galactic center a maximum likelihood method (binned and unbinned) with likelihood ratio test statistics was exercised on the SCdata2. The method uses both the spectral models of the Dark Matter and the astrophysical sources close to the galactic center and the spatial distributions of these sources.  
The energy spectra and spatial distributions of the sources can be modeled using parameterizations, tables or FITS maps. We will have the possibility to fit the mass, normalization and annihilation branching fractions~\cite{Jeltema:2008hf}.

We are now in the process of applying this procedure to Fermi LAT data. Claims of discovery or sound constraints will require thorough understanding of the complicated galactic center region, in particular the diffuse galactic model will be especially tuned for this region by the Fermi LAT collaboration.

\section{Comparison to particle physics models}

In the preceding section we have considered a generic model for Dark Matter that causes the
$\gamma$-ray signal. We only assumed a WIMP that gives a mono-energetic quark (or fermion)
spectrum. 
In this context the cross-section and mass of the WIMP are
free parameters. However, there are several extensions to the Standard Model of Particle
Physics which predict a particle state which could constitute the WIMP. The
most studied class of such models is supersymmetry, in particular minimal supersymmetric
extensions to the Standard Model (MSSM) and its constrained version mSUGRA, in which
the soft supersymmetry breaking terms derive from a high-energy supergravity theory
with a common scalar mass $m_0$ and a common gaugino mass $m_{\nicefrac{1}{2}}$ at the 
GUT scale (for details see \cite{Hall:1983iz,Ohta:1982wn}). 

\begin{figure}[h]
  \includegraphics[width=.48\textwidth, viewport= 10 8 878 827,clip]{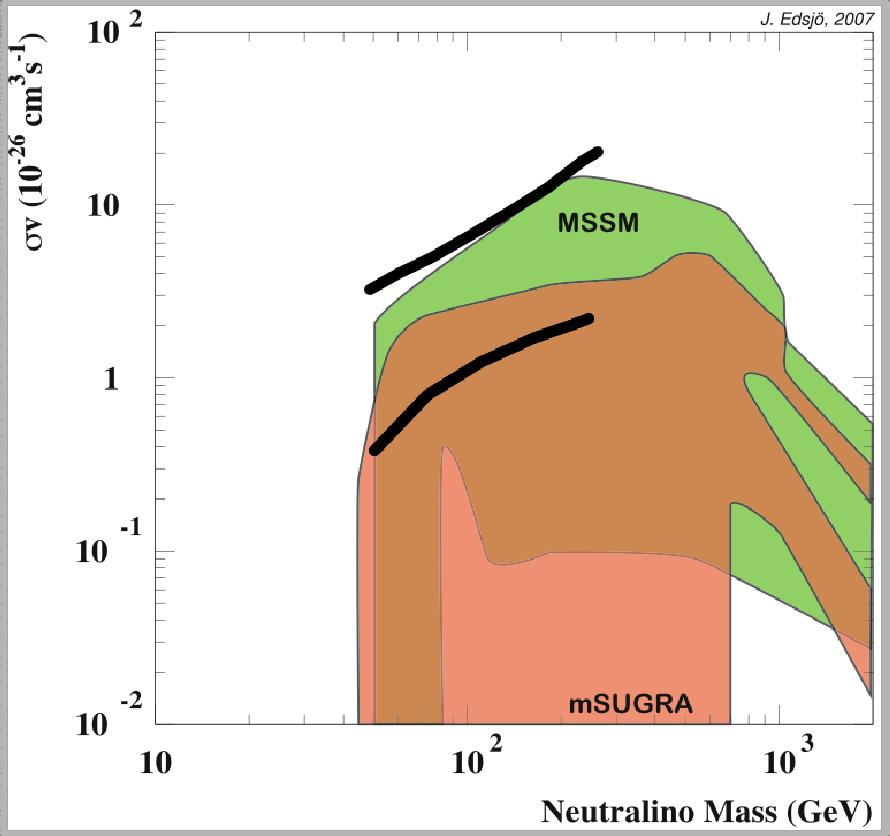}
  \caption{MSSM and mSUGRA models in the $(<\sigma v >, m_\chi)$  plane. The dark lines indicate the Fermi \unit[5]{$\sigma$} sensitivity for a Dark Matter search in the galactic center (upper line) and in the galactic halo (lower line)~\cite{Baltz:2008wd}. \label{models}}
\end{figure}

In Fig~\ref{models} we show the set of mSUGRA and MSSM models which pass all accelerator constraints and are consistent with WMAP data in the
parameters in which we calculated the LAT sensitivity, i.e. the $(<\sigma v >, m_\chi)$ 
plane. The Fermi sensitivity for a Dark Matter search in the galactic halo covers partly  the 
mSUGRA phasespace as well as the MSSM phasespace. For a Dark Matter search in the galactic center the Fermi sensitivity touches at least the border of the MSSM phasespace.

\section{Conclusions and Outlook}

In these proceedings we summarize the searches for particle Dark Matter to be performed with the Fermi LAT instrument. The Fermi LAT collaboration pursues complimentary searches for Dark Matter signal, each presenting its own advantages and challenges. 
In this contribution we discussed in detail the ongoing analysis on the Dark Matter search in the galactic center region as well as the coverage of the Fermi LAT sensitivity for this analysis in the parameter space of the MSSM and mSUGRA models.

Fermi is in routine science operations since August 11th, 2008. 
The establishment of a detection will require thorough understanding of the involved backgrounds and possibly joint observations with the ground based Cherenkov telescope experiments H.E.S.S., MAGIC, VERITAS and CANGAROO.

\begin{theacknowledgments}
I would like to thank all members of the Dark Matter and New Physics Working Group of the Fermi LAT who contributed to this note.
\end{theacknowledgments}

\bibliographystyle{aipproc}   

\bibliographystyle{aipproc}   

\end{document}